\documentclass[aps,showpacs,nofootinbib,superscriptaddress]{revtex4}

\usepackage{graphicx}
\usepackage{dcolumn}

\newcommand{\reales}{{\rm R}\hspace{-1ex}\rule{0.1mm}{1.5ex}\hspace{1ex}}
\begin{document}

\title{Renormalization of the $^1S_0$ One-Pion-Exchange NN Interaction
       in Presence of Derivative Contact Interactions} \author{
       J. Nieves} \affiliation{Departamento de F\'{\i}sica Moderna,
       Universidad de Granada, E-18071 Granada, Spain.}

\begin{abstract} 
\rule{0ex}{3ex} We use standard distorted wave theory techniques and
dimensional regularization to find out solutions of the
nucleon-nucleon Lippman--Schwinger equation with a potential that
includes one--pion exchange and additional contact terms with 
derivatives. Though for simplicity, we restrict the discussion to the
$^1S_0$ channel and to contact terms containing up to two derivatives,
the generalization to higher waves and/or number of derivatives is
straightforward. The undetermined low energy constants emerging out of
the renormalization procedure are fitted to data.
\end{abstract}


\maketitle

%=========================================

 \section{Introduction}

Over the last years, Effective Field Theory (EFT) methods have become
the standard tool to deal with strong interactions in the
non-perturbative regime. In the form of Chiral Perturbation Theory
(ChPT), they have been used with some success in both the mesonic and
single nucleon sectors. The EFT of nuclear forces based on a chiral
expansion was originally suggested by Weinberg~\cite{We91}, and since
then a lot of work has been devoted to gain a better understanding of
the two-nucleon interaction at low and intermediate
energies~\cite{OK92}--\cite{Ol02}. While at the beginning, these
studies did not  aim at substituting the highly successful {\it
realistic} potentials built from meson exchanges (Bonn-J\"ulich,
Nijmegen, Argonne, $\cdots$ potentials), the importance of uncovering
such an EFT cannot be ignored, since this theory will allow for
rigorous calculations of both elastic and inelastic processes in
systems with two or more nucleons, in a framework consistent with the
Standard Model of strong and electroweak interactions. Furthermore,
the latter EFT works provide an accurate description of nucleon-nucleon
(NN) phase-shifts for several partial waves and in a wide range of
energies~\cite{EM02}.  Weinberg's original proposal was to determine
the NN potentials using the organizational principles of ChPT and then
to insert these potentials into the Lippman-Schwinger Equation (LSE)
to solve for NN scattering amplitudes. Though such a scheme has proved
to be successful~\cite{OK92}, \cite{PMR95}, \cite{EGM98}, it suffers
from formal inconsistencies, in particular, divergences that arise at
a given order in the chiral expansion cannot be absorbed by terms of
the same order in the Lagrangian~\cite{KSW96},\cite{KSW98}.  The
trouble arises, since Weinberg's power counting assumes that the
coefficients of the higher order operators were set by a {\it typical}
Quantum Chromo Dynamics (QCD) scale, but in practice it turns out that
there appear explicit $M$ factors (being $M$ the nucleon mass) in the
needed counter-terms to renormalize the amplitudes, which formally
decrease the degree of the operator in the Weinberg's counting. This
was partially solved by Kaplan, Savage and Wise (KSW) who introduced a
new scheme where pions are treated
perturbatively~\cite{KSW98}. However, it turns out that the KSW
expansion converges slowly in the $^1S_0$ ($^{2I+1}L_J$) channel and
does not converge at all in the $^3S_1-$$^3D_1$ coupled
channels~\cite{TH98},~\cite{FMS00}. Very recently, a so called {\it
expansion about the chiral limit} has been proposed in
Ref.~\cite{Be01}, which seems to be equivalent to KSW power counting
in the $^1S_0$ channel and to Weinberg power counting in the
$^3S_1-$$^3D_1$ coupled channels. 

In this work we adopt the following point of view. For a scattering
process involving external momenta $< q$, one should only consider a
Lagrangian/potential which explicitly includes light degrees of
freedom for which $m < q$. The effects of heavy virtual particles
appear as an infinite number of non-renormalizable operators
suppressed by some mass scale relevant to the degrees of freedom
excluded from the theory.  At low-intermediate momenta, below twice
the pion mass, it should be enough to include explicitly
one-pion-exchange, and simulate the rest of the physical contributions
by a tower on non-renormalizable contact 
interactions, organized as a derivative expansion. Given a potential,
which includes contact terms up to some order, the LSE performs  an
infinite sum of diagrams, which in general would require to be
renormalized. After renormalization the coefficients of the contact terms
would contain, in general, contributions from all orders in $m_\pi$
(pion mass).

Most of the formal problems appearing within the Weinberg's scheme are
linked to the use of a somehow restrictive scheme to renormalize the
LSE~\cite{GJ01}--\cite{ES01}. Such a restricted conception of the
renormalization of the LSE would also lead to unexpected consequences
in other scenarios, as for example $\pi\pi$ scattering in the
$\rho-$channel~\cite{ej99}. Among the abundant literature on the
subject, the framework presented in Ref.~\cite{FTT99} constitutes a
first step towards the renormalization scheme proposed in this work.
There, a subtracted LSE is derived, and the numerical values of the
fitted Low Energies Constants (LEC's) are not translated to values for
any Ultraviolet (UV) cutoff.

The aim of this paper is to present a
renormalization scheme of the LSE for NN scattering by a combination
of known long-range and unknown short-range potentials. 
For simplicity, we restrict the discussion
to the $^1S_0$ channel, where we show that the proposed scheme
describes phase-shifts up to Center of Mass (CM) nucleon momenta of
the order of 260 MeV. This renormalization scheme is based on previous
findings on $\pi\pi$~\cite{ej99} and meson-baryon
scattering~\cite{ej01}--\cite{OR98}, it leads to renormalized
amplitudes which fulfill exact elastic unitarity, and it is easily
generalized to study higher waves and/or number of derivatives in the
contact part of the interaction.

\section{Effective potentials and the LSE} After projecting into the 
$^1S_0$ partial wave, the NN LSE for a CM nucleon kinetic energy $E$, reads:
\begin{equation}
T(E; p,p^\prime) = V(p,p^\prime) + \int^{+\infty}_0 dq
q^2 \frac{V(p,q)T(E;q,p^\prime)}{2m E-q^2 +{\rm i}\epsilon} \label{eq:lsedef}
\end{equation}
with $m=469.46$ MeV the NN reduced mass, for which we take that of the
neutron-proton system, $p$ and $p^\prime$ are the initial and final
relative momenta of the two nucleons, and $V(p,p^\prime)$ the $^1S_0$
NN potential. The normalization of the scattering amplitude $T(E;
p,p^\prime)$ is such that on the mass shell,
$p=p^\prime=\sqrt{2mE}\equiv k$, it is related to the phase shifts,
$\delta$, by:
\begin{equation}
T(k) =  -\frac{2}{\pi}\frac{e^{2{\rm i}\delta(k)}-1}{2{\rm i}k} 
\end{equation}
The potential consists of the one-pion-exchange
potential,$V_\pi$, plus contact terms,$V_s$, with up to two
derivatives~\cite{KSW96}: 
\begin{eqnarray}
V(p,p^\prime)&=& V_s(p,p^\prime)+V_\pi(p,p^\prime)\nonumber\\
V_s(p,p^\prime)&=& 
 g_0 +g_1 (p^2+p^{\prime\, 2}) \nonumber \\
V_\pi(p,p^\prime) &=& -\frac{2m \alpha_\pi}{\pi}  \int_{-1}^{+1}
\frac{dx}{p^2+p^{\prime\,^2}-2pp^\prime x+m_\pi^2}
 = \frac{m \alpha_\pi}{\pi} \frac{1}{pp^\prime}
\log\frac{p_-^2+m_\pi^2}{p_+^2+m_\pi^2}, \qquad {\rm with} \quad  
\alpha_\pi = \frac{g_A^2 m_\pi^2}{16\pi f_\pi^2} \label{eq:potentials}
\end{eqnarray}
with $p_{\pm} = p \pm p^\prime$, $m_\pi=138$ MeV and $f_\pi=93$ MeV
the pion mass and weak decay constant and finally $g_A =1.25$ the
axial coupling constant. Note that the contact term includes the
$\delta^3(\vec{r}~)$ contribution from one pion exchange, and it also
does the leading and sub-leading contributions in the derivative
expansion of all shorter distance effects, such as Two Pion Exchange
(TPE)\footnote{Note that, the full structure of the logarithmic terms,
stemming from the pion loops implicit in the TPE contribution, cannot
be entirely accounted for the contact terms. Indeed, for CM transferred 
momenta above
$2 m_\pi$, TPE contributions would have to be treated in the same
footing as the one-pion exchange ones.}, intermediate $\Delta$'s,
$\omega$ exchange, etc $\cdots$. Such a procedure suffers from some
limitations, for instance, since  the TPE potential has not been
explicitly included one cannot relate NN scattering to other
processes as pion-nucleon, pion-deuterium, etc \ldots scattering. The
LEC's $g_0$ and $g_1$ are not fixed by chiral symmetry and have to be
determined by a fit to the phase shifts, as we will see below.
Scattering by short-range interactions in the presence of a known
long-range potential,$V_\pi$, can be treated by Distorted Wave Theory
(DWT)~\cite{BB02}. We write the full scattering matrix
as\footnote{Since the angular integrals have been already performed
when projecting into the $^1S_0$ wave and taking into account that do
not lead to UV divergences, in what follows we will assume that the
measure in momentum space and the free nucleon propagator are given by
$\int_0^{+\infty} dq q^2$ and $G_0(E;p,p^\prime) =
\frac{\delta(p-p^\prime)}{p^2} \frac{1}{2m E-p^2+{\rm i}\epsilon}$,
respectively.}
\begin{equation}
  T = T_\pi + (I+T_\pi G_0) \hat{T}_s (I+G_0 T_\pi) \label{eq:dwt}
\end{equation}
with $I$ the identity, $G_0$ the free nucleon Green
function and $T_\pi$ the scattering matrix for $V_\pi$ alone. Besides, 
$\hat{T}_s$ describes the scattering between distorted waves of
$V_\pi$. It satisfies the LSE
\begin{equation} 
\hat{T}_s = V_s + V_s G_\pi \hat{T}_s 
\end{equation}
where $G_\pi$ is the nucleon Green function in the presence of
$V_\pi$, i.e., $G_\pi = G_0+G_0 T_\pi G_0$. To solve the LSE of
Eq.~(\ref{eq:dwt}), the full off-shell scattering matrix $T_\pi(E;
p,p^\prime)$ is required, which is determined by solving the LSE of
Eq.~(\ref{eq:lsedef}) with the obvious substitution of $V \to
V_\pi$. This equation is finite (it corresponds to the scattering by
the usual Yukawa force studied in most books of Quantum Mechanics) and
does not require to be renormalized, being then possible a
numerical evaluation. We have obtained $T_\pi$  by discretizing the momentum
space and using the inverse matrix algorithm. Some
results are shown in the top panel of Fig.~\ref{fig:1s0}. To
obtain $\hat{T}_s(E; p,p^\prime)$, we solve
\begin{eqnarray}
\hat{T}_s(E;p,p^\prime) &=& V_s(p,p^\prime) +  \int^{+\infty}_0 dq
q^2 \frac{V_s(p,q)\hat{T_s}(E;q,p^\prime)}{k^2-q^2+{\rm
i}\epsilon}+ \int^{+\infty}_0 dqdq^\prime \frac{q^2 q^{\prime
\,^2} V_s(p,q)T_\pi(E;q,q^\prime){\hat T}_s(E;q^\prime,p^\prime)}
{\left (k^2-q^2+{\rm
i}\epsilon\right)\left (k^2-q^{\prime \, 2} +{\rm
i}\epsilon\right)}
\end{eqnarray}
The above equation can be reduced to a linear algebraic system of
equations, which solution is straightforward, 
\begin{equation}
\hat{T}_s(E,p,p^\prime)
= \alpha + \beta(p^2+p^{\prime\, 2})+ \gamma p^2 p^{\prime\,
2},\label{eq:short}
\end{equation} 
with the energy dependent functions 
\begin{eqnarray}
\alpha &=& \frac{g_0+g_1^2 K_4}{\Delta}, \qquad \qquad 
\beta = \frac{g_1-g_1^2 K_2}{\Delta}, \qquad \qquad 
 \gamma = \frac{g_1^2K_0}{\Delta}
\nonumber \\ 
\Delta &=& 1 -g_0 K_0 - 2g_1 K_2 + g_1^2 \left (K_2^2-K_0K_4
\right). 
\end{eqnarray}
The UV divergent integrals $K_n(E) = I_n(E) + J_n(E)$ 
are defined by:
\begin{eqnarray}
I_n(E) &=& \int^{\infty}_0 dq \frac{q^{n+2}}{k^2-q^2+{\rm i}\epsilon}\\
&&\nonumber\\
J_n(E) &=& \int^{\infty}_0 
\frac{dqdq^\prime q^a q^{\prime \,b} T_\pi(E;q,q^\prime)}{\left(k^2-q^2+{\rm
i}\epsilon\right)\left(k^2-q^{\prime \,2}+{\rm i}\epsilon\right)}
\end{eqnarray}
and the pair of integers $(a,b)$ take the values (2,2),(4,2) and (4,4)
for the $n=0,2$ and 4 cases respectively.  

 We use the Dimensional
Regularization (DR) scheme, since it preserves chiral and gauge symmetry
and Galilean invariance, which makes the integrals relatively simple
to evaluate. DR discards all power-law divergences of the type $\int
dq q^n$, which makes finite all the $I_n$ integrals define above,
\begin{equation}
I_n(E) = k^n \times  {\rm i} \left ( -\frac{\pi k}{2}
\right )
\end{equation}
To simplify the $J_n$ integrals it is useful to realize that in DR the 
linearly UV divergent integral $\int_0^{+\infty} dq q^2 V_\pi(q,p)$ is
finite ($= 2m \alpha_\pi m_\pi$) and independent of
$p$~\cite{KSW96}. Making use of the LSE satisfied by $T_\pi$ to get
expressions involving the above integral, the $J_2$  and $J_4$
integrals can be related to the $J_0$ one,
\begin{eqnarray}
\frac{J_n (E)}{2m} = (E-m_\pi\alpha_\pi)J_{n-2}(E) + {\rm i}
\frac{\pi}{2}m_\pi \alpha_\pi k^{n-1} \label{eq:defj}
\end{eqnarray}
for $n=2,4$. Besides, $J_0(E)$ is logarithmically divergent and it only
requires one subtraction, i.e., ${\bar J}_0 (E) = J_0(E)-J_0(0)$ is
finite and can be numerically evaluated. Plugging Eq.~(\ref{eq:short})
into Eq.~(\ref{eq:dwt}) one gets for the on-shell scattering amplitude
$T(E;k,k)\equiv T(k)$ :
\begin{eqnarray}
T(k) &=& T_\pi(k)+ {\hat T}_s(k) + 2 \left \{(\alpha + \beta k^2) L_0
+ (\beta+\gamma k^2) L_2 \right \} 
+ \alpha L_0^2 + 2\beta
L_0L_2+\gamma L_2^2 
\end{eqnarray}
with $T_\pi(k)$ and $ {\hat T}_s(k)$ the long and short range on-shell
matrices and 
\begin{equation}
L_n(k) = \int^{\infty}_0 
\frac{dq q^{n+2} T_\pi(E;k,q)}{\left(k^2-q^2+{\rm
i}\epsilon\right)}
\end{equation}
The integral $L_0$ is finite and the UV divergent integral $L_2$ in DR becomes
finite, i.e., $L_2(k) = 2m \times $ $\left \{ (E-\alpha_\pi m_\pi)L_0(k)
- \alpha_\pi m_\pi \right \}$.  With all above results one gets 
\begin{eqnarray}
T(k) &=& T_\pi (k) + \left (1 + L_0(k)\right)^2 {\hat T}_s(E;{\hat
k},{\hat k}) =  T_\pi (k) + \frac{\left (1 +
L_0(k)\right)^2}{V_s^{-1}(\hat{k},\hat{k})+ {\rm i}\pi k/2 - {\bar
J}_0 (E) -J_0(0) }
\end{eqnarray}
with ${\hat k}^2 = k^2 - 2m\alpha_\pi m_\pi$. Elastic unitarity is
exactly fulfilled thanks to a Watson's type relation
satisfied  by $L_0(k)$\footnote{This relation is  easily  obtained from the
Optical Theorem satisfied by $T_\pi$, i.e., $2 {\rm Im}
T_\pi(E;p_1,p_2) 
= -\pi k T_\pi(E;p_1,k) T_\pi^*(E;p_2,k)$.},
\begin{equation} 
L_0(k) = \left (l_0(k) - {\rm i} \pi
k/2\right )T_\pi(k), \quad l_0 \in \reales \label{eq:defl0}
\end{equation} 
and that 
\begin{equation} 
{\rm Im} {\bar J}_0 (k) = -\left \{ (\pi
k/2)^2 -l_0(l_0+2{\rm Re}T_\pi^{-1}(k))\right \} {\rm Im} T_\pi (k).
\end{equation} 
Indeed, the on shell scattering matrix, $T$, can be re-written in a
form where unitarity is manifest, 
\begin{eqnarray}
T^{-1}(k) &=& T^{-1}_\pi(k) - \frac{\left (l_0(k)+{\rm
Re}T_\pi^{-1}(k)\right )^2}{ {\rm Re}T_\pi^{-1}(k) + g(k)} \label{eq:t_res}\\
g(k) &=& V_s^{-1}(\hat{k},\hat{k}) -{\rm
Re}J_0(k)+ \frac{{\rm Im}{\bar J}_0(k)}{\tan\delta_\pi(k)}
\end{eqnarray}
Above threshold, the function $g(k)$ is  real, and
$\delta_\pi(k)$ are the phase shifts deduced from $T_\pi(k)$. The
functions $l_0(k)$ and ${\rm Re}{\bar J}_0(k)$ and the phase shifts
$\delta_\pi$ are plotted in the top panel
of Fig.~\ref{fig:1s0}. Note that in the complex $E$ plane, $g(k)$ does
not have right-hand-cut, but it does have a left-hand-cut due to pion
exchange. Before going ahead we should study how to get rid of the UV
divergences.

\section{Renormalization Scheme.}
The DR scheme has led to a drastic reduction of UV divergences and
thus we are just left with the logarithmic divergent
integral\footnote{The divergent part of $J_0(0)$ is given by the
integral 
\begin{equation}
\int_0^{+\infty} dqdpV_\pi(q,p) = -2m \alpha_\pi \int_0^{+\infty}
\frac{dq}{q} \left (
\frac{\pi}{2} -{\rm arctan}\frac{m_\pi}{q} \right ).
\end{equation}
}:$J_0(0)$, which should be absorbed into a redefinition of the
coupling constants of the potential, provided one included in the
potential all terms consistent with the symmetry principles. However,
the amplitude cannot be made finite by simply redefining the couplings
$g_0$ and $g_1$. This problem arises because we have not included
operators with more than two derivatives in the potential, which are
needed as counter-terms to render multiple insertions of the two
derivative operator finite. Indeed, the divergent constant $J_0(0)$
appears in the function $g(k)$ defined above, and it seems natural to
define a renormalized contact potential ${V^R_s}^{-1}({\hat k},{\hat
k}) = V_s^{-1}({\hat k},{\hat k}) -J_0(0)$ which leads to an infinite
series of powers of $k^2$ for $V_s^R$.  Since the LSE performs a
non-perturbative resummation of diagrams, in principle one should also
add a series of counter-terms to the bare amplitude such that the sum
of both becomes finite. At each order in the $k^2$ (or derivative)
expansion, the divergent part of the counter-term series is completely
determined. However, the finite part remains arbitrary. It means that
the coefficients of each of the terms in the $k^2$ series, implicit in
the definition of ${V^R_s}^{-1}$, become undetermined and have to be
fitted to data or, if possible, evaluated in QCD. Thus, the scenario
might look like pessimistic, and because of the appearance of
divergences, $g(k)$ turns out to be a completely undetermined
function. This situation has some resemblances to ChPT in the
meson-meson and meson-baryon sectors. For simplicity, let us pay
attention to elastic $SU(2)$ $\pi\pi$ scattering. It is well
known~\cite{GL84} that the divergences arising at one loop, ${\cal
O}(p^4)$, cannot be absorbed into a redefinition of the leading
terms\footnote{Power divergences can be absorbed into a redefinition
of the leading terms, but logarithmic ones cannot, and require the
inclusion of higher order structures.}, ${\cal O}(p^2)$, of the
Lagrangian. New counter-terms, $l_i$, higher in the chiral expansion
are needed to absorb the divergences, which finite parts $ {\bar
l}_i$, remains undetermined, cannot be fixed by chiral symmetry and
have to be fitted to data. These LEC's are fundamental parameters of
the EFT, which contain the contribution at low energies of higher
degrees of freedom, which have been integrated out. The evaluation of
the divergent parts of the amplitudes with an UV cut-off does not
necessarily provide a reasonable estimate for them\footnote{If an UV
cut-off $\Lambda$ is employed, the ${\cal O}(p^4)$ contributions in
all isospin-angular momentum elastic $\pi\pi$ scattering channels will
be determined just by one parameter, $\Lambda$, while at this order
there are four independent parameters, $ {\bar l}_{1,2,3,4} $, which,
for instance, incorporate the effect of the $\rho$ and other
resonances in the amplitudes (see for instance discussion in Section 3
(pages 63--70) of second entry of Ref.~\cite{ej99}).}. To be
predictive, one can adopt here a renormalization scheme, similar to
that used in Refs.~\cite{ej99}--\cite{mb02} to renormalize the Bethe
Salpeter equation for meson-meson and meson-baryon scattering, which
reduces the enormous freedom discussed above.  Since for small
momenta, higher derivative operators should have a tiny influence, we
choose to reduce all this proliferation of LEC's, by imposing
relations, among all LEC's associated to counter-terms with a number
of derivatives higher than the terms included in the potential, in
such a way that the renormalized amplitude can be cast, again, as in
Eq.~(\ref{eq:t_res}) and therefore it fulfills elastic unitarity. This amounts
in practice, to interpret the previously divergent quantity $J_0(0)$
as a renormalized undetermined parameter. After having renormalized,
we add a superscript $R$ to differentiate between the previously
divergent, $J_{0}(0)$, and now finite quantity\footnote{ Thus, in the
calculation presented in this work, we relate the finite parts of the
$k^6$, $k^8$, etc..., contributions to that of the $k^4$ one (note
that, the $k^4$ contribution shows up at the one-loop level induced by
contact terms quadratic in momenta: $p^2$ and $p'^2$), which is
determined by $J^R_0(0)$. The relation is such that elastic unitarity
is restored, and thus this scheme differs from those where the higher
order terms are set to zero.  Indeed, setting to zero the higher order
terms is as arbitrary as setting them to any other value. At next
order in this derivative expansion, it is to say when terms of the
type $p^2~p'^2$, $p^4$ or $p'^4$ are included explicitly in the
potential, the LSE would generate terms of order $k^6$ and higher in
the amplitude. Thus, the expansion based on this renormalization
scheme is systematically improvable.}, $J^R_{0}(0)$. This parameter
and therefore the renormalized amplitude can be expressed in terms of
physical (measurable) magnitudes. The estimate given by means of a
{\it reasonable} UV cut-off for the numerical value of $J^R_{0}(0)$
might not be good. For instance, Eqs.~(A15) of second entry of
Ref.~\cite{ej99} illustrate this point, if an UV cutoff is employed,
the divergent integrals appearing there will be independent of the
$IJ$ channel, and to get a reasonably value for $I^{R,I=1}_0$
unrealistic scales or cut-offs of the order of 300 GeV will be
needed. This is due to the special nature of the $\rho$-resonance, and
that is not the case, for instance, for the $s-$waves.

Thus, we are proposing an expansion of the short range part of the
potential in powers of $k^2$, i.e., delta function and its derivatives
in coordinate space, and for a given potential, to compute $T$ to all
orders in the $k^2$ expansion to restore exact unitarity. Thus, the
contact terms effectively account for degrees of freedom higher than
the pion and one should expect the scheme to work up to the first
threshold, likely two pion production ($k_{LAB}\approx 2m_\pi$), which is
around the normal nuclear matter Fermi momentum. A word of caution
must be said here; this expansion is not equivalent to a chiral
expansion in $m_\pi$, since the coefficients of the contact
interaction would contain contributions from all orders in $m_\pi$, as
it was firstly pointed out in Ref.~\cite{KSW96}.

\section{Results and Concluding Remarks}
\begin{figure}
\centerline{\includegraphics[height=13.8cm]{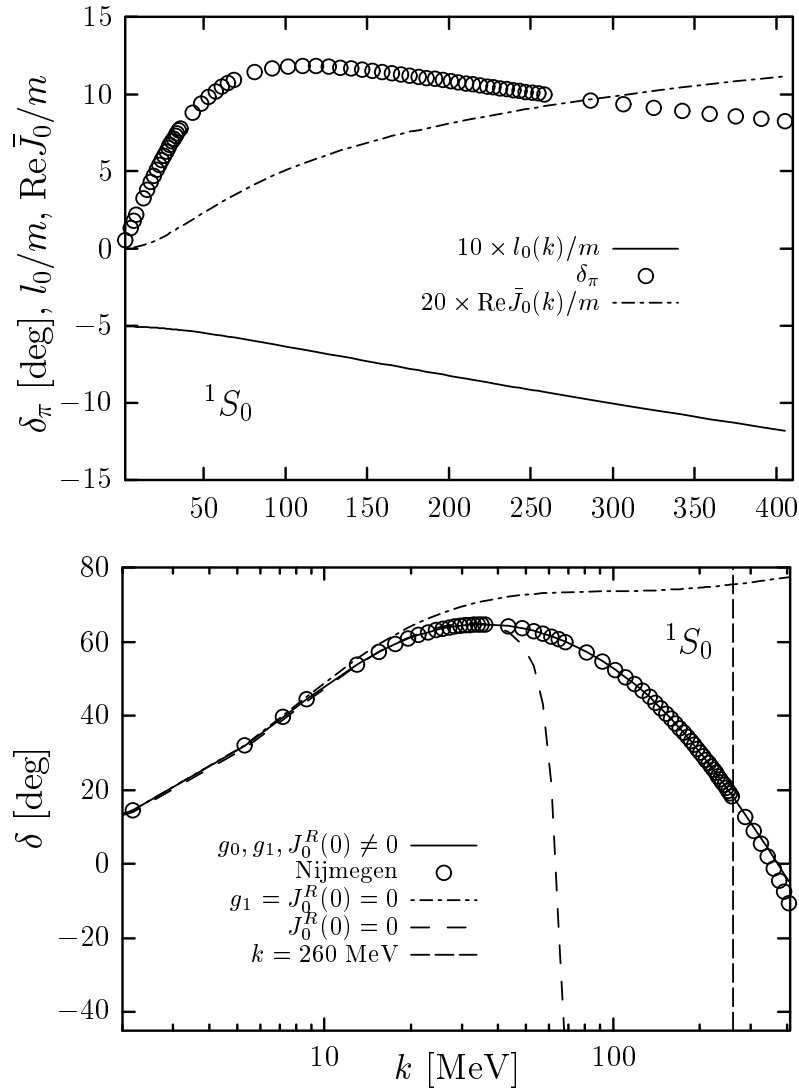}}
\vspace{0.1cm}
\caption[pepe]{\footnotesize Top panel: Several quantities extracted
from $T_\pi$: phase shifts and functions $l_0(k)$ and ${\rm Re}{\bar
J}_0(k) = {\rm Re}J_0(k)-J_0(0)$ defined in
Eqs.~(\protect\ref{eq:defj}) and~(\protect\ref{eq:defl0}),
respectively. From the energy dependence of $\delta_\pi(k)$, we deduce
the effective range parameters (see Eq.~(\protect\ref{eq:ere})) $a_\pi
= -0.88$ fm and $r_{0\pi}=12.38$ fm. Bottom panel: $^1S_0$ $np$ phase
shifts in degrees plotted versus CM momentum (log scale). Solid line
stands for the results of this work (Eq.~(\protect\ref{eq:fit})),
while other two approaches with ${ J}^R_0(0)= 0$ and ${
J}^R_0(0) = g_1 = 0$ respectively, 
are also shown. Data are taken from the phase-shift 
analysis of Ref.~\protect\cite{NJ93}, and data for momenta above the
vertical line (260 MeV) have not been included in the fit.}
\label{fig:1s0}
\end{figure}
After the above discussion, it is clear we have three undetermined
LEC's: $g_0$, $g_1$ and $J_0^R(0)$, which we obtain from a $\chi^2-$
fit to Nijmegen phase-shifts~\cite{NJ93}.  To perform the fits, we
assume errors on the phase shifts, $\delta$ (in degrees), given by
$0.5 + {\rm abs} (\delta)\times {\rm max}\left(0.1, E_{Lab}[{\rm
MeV}]\right)/100$, where $E_{Lab}$ is the kinetic energy in the
laboratory frame. We fit data from threshold up to a certain value,
$k_{\rm max}$, of the CM nucleon momentum. We determine $k_{\rm max}$,
by studying the dependence of $\chi^2/dof$ on it, and fix it in the
region of 260 MeV, since the inclusion of simply a new datum of higher
energy would double the value of $\chi^2/dof$. Thus, we have fitted 62
phase shift values for which $E_{Lab}$ has varied from 0.01 to 142 MeV, and
results are shown (solid line) in the bottom panel of
Fig.~\ref{fig:1s0}. Best fit parameters are 
\begin{equation}
g_0 m = -0.276 (6), \qquad \qquad
g_1 m^3 = 0.347 (14), \qquad \qquad
\frac{J_0^R(0)}{m} = -3.21 (8),   \label{eq:fit}
\end{equation}
 with $\chi^2/dof = 0.016$. In the above equation, statistical errors
have been shown in brackets, affect to the last digit of the
parameters and are given by the square root of the diagonal elements
of the covariance matrix, $v_{ij} =
\left[\left(\frac{1}{2}\frac{\partial \chi^2}{\partial b_k\partial
b_l}\right)^{-1}\right]_{ij}$, with $b_i$ any of the three fitted
parameters. It is worth to mention that from the above matrix, one
also learns that the pairs $g_0, g_1$ and $g_0, J_0^R(0)$ are highly
anticorrelated, with correlation coefficients
($r_{ij}=v_{ij}/\sqrt{v_{ii} v_{jj}}$) smaller than $-0.99$.  This
means that effectively there is just one independent parameter. We
 will come back to this point below.

If the derivative terms of the contact interaction are set to zero,
$g_1=0$, the UV divergence can be absorbed in $g_0$~\cite{KSW96} and
thus one is just left with one parameter, which can be fitted to
data. Yet, though $g_1 \ne 0$, one can arbitrarily set the renormalized
coefficients of the higher order terms to zero, i.e., take ${ 
J}_0^R(0) = 0$, and fit the non-zero LEC's, $g_0$ and $g_1$, to
data. Both procedures are also shown in Fig.~\ref{fig:1s0}, the first
one leads to the scheme developed in Ref.~\cite{FTT99}
and the second one was studied for the very first time in 
Ref.~\cite{KSW96}. As seen in the figure, these two schemes do not
work for momenta higher than 10 and 40 MeV respectively, while the
solid line provides a good description of data up to momenta of the
order of 260 MeV. Indeed, from Eq.~(\ref{eq:fit}) we have 
$\sqrt{\left |g_0/2g_1\right |}\approx  295$ MeV, while such ratio
takes a value around of 70 MeV when $J^R_0(0)$ is set to zero. 

The scattering length, $a$, and the effective range, $r_0$, are
defined from the effective range expansion,
\begin{eqnarray}
 T^{-1}(k)= -\frac{\pi}{2} \left (-\frac{1}{a} + \frac12 r_0 k^2 +
 \cdots - {\rm i} k \right ) \label{eq:ere}
\end{eqnarray}
Fitting the inverse of our amplitude, up to $k=48.5$ MeV, to the above
formula, we get $a=-23.65$ fm and $r_0= 2.63$ fm, in reasonable
agreement with the Nijmegen results~\cite{NJ93} ($-23.7$ fm and 2.73 fm,
respectively) also obtained in this way.

To finish, we would like just to summarize the results obtained in
this work. We advocate for an expansion of the short range part of the
potential in powers of $k^2$ and  computing $T$ to all
orders in the $k^2$ expansion to restore exact elastic unitarity. 
DWT techniques and DR have been used to solve the $^1 S_0$ NN
LSE with a potential,
which consists of pion exchange and contact terms with up to two
derivatives. The procedure of solving the LSE is quite simple and an
explicit expression, where exact elastic unitarity is manifest, has
been given as well (see Eq.~(\ref{eq:t_res})). A particular
renormalization scheme has also been discussed. It is based on
previous findings on meson-meson and meson-baryon systems and its main
ingredient is to realize that an EFT is not a renormalizable field
theory in the sense of QCD, i.e., with a finite number of
counter-terms. Thus, to keep finite the amplitude, obtained after
performing the non-perturbative resummation implicit in the LSE, would
require, in general, the addition of an infinite set of counter-terms
in the short distance potential ($k^2$ series). The finite parts of
the coefficients of the series would remain undetermined (LEC's) and
encode the contribution of higher degrees of freedom, not explicitly
included. An UV cut-off can effectively account for this freedom only
up to some momentum scale, which will  depend on the terms
explicitly included in the potential\footnote{The EFT only becomes
cut-off independent when all counter-terms compatible with the
underlying symmetry are included in the potential.}. For a contact
potential including up to $k^2$ terms, we find a good description of
phase shifts up to CM NN momenta of the order of 260 MeV with the
inclusion of just one additional parameter, $J^R_0(0)$. 

In Ref.~\cite{EGM98}, an UV cut-off is used to regularize the
 amplitudes, and with reasonable values of it, in the range $0.6-1$
 GeV, a good description of data is found. Note, however, that
 this does not have always to be the case and depends, as we mentioned
 above, of the physical system and of the order of the expansion
 included in the potential.  Thus, to describe the $\rho-$resonance in
 $\pi\pi$ scattering, one is left to deal with UV cut-off's of the
 order of 300 GeV~\cite{ej99},\cite{O97}. Unrealistic values of the UV
 cutoff are also needed to account for the $N(1650)-$resonance in
 meson baryon scattering~\cite{ej01},\cite{mb02}, but this is not the
 case, for instance, for the $\sigma$ and $f_0(980)$ resonances in $\pi\pi$
 scattering~\cite{O97} or the $\Lambda (1405)$ resonance in meson baryon
 scattering~\cite{OR98}.

Besides, we find extremely big
anticorrelations between $J^R_0(0)$ and the coefficients, $g_0$ and $g_1$, of
the iterated short distance potential and among the two later ones, as
well. This indicates that the higher derivative operators in the EFT,
generated in this scheme by the inclusion of $J^R_0(0)$, are actually
highly correlated. This might support the idea that though, the higher
derivative operators are controlled by a scale that diverges as $|a|
\to \infty$, thanks to these high correlations, the effects that
diverge with $a$ cancel, as it was pointed out in Ref.~\cite{KSW96}.

To include more derivatives is trivial and would result, besides the
left hand cut due to pion exchange and accounted for by means of
$T_\pi$, in a higher order {\it pad\`e} approximant for the function
$g(k)$ in Eq.~(\ref{eq:t_res}). The extension of the procedure to
higher partial waves is also straightforward although cumbersome and
will be presented elsewhere~\cite{jn03}.

\begin{acknowledgments}

I warmly thank to E.Ruiz-Arriola for useful discussions.  This
research was supported by DGI and FEDER funds, under contract
BFM2002-03218 and by the Junta de Andaluc\'\i a.

\end{acknowledgments}


\begin{thebibliography}{99}

\bibitem{We91} S. Weinberg, Phys. Lett. {\bf B251} (1990) 288; 
Nucl. Phys. {\bf B363} (1991)2; Phys. Lett. {\bf B295} (1992) 114.
\bibitem{OK92} C. Ord\'o\~nez and U. van Kolck, Phys. Lett. {\bf B291}
(1992) 459;  U. van Kolck, Phys. Rev. {\bf C49}
(1994)  2932; Prog. Part. Nucl. Phys. {\bf 43} (1999)
337; C. Ord\'o\~nez, L. Ray and U. van Kolck,
Phys. Rev. Lett. {\bf 72} (1994) 459; Phys. Rev. {\bf C53}
(1996)  2086.  
\bibitem{PMR95} T.Park, D.Min and M. Rho, Phys. Rev. Lett. {\bf 74}
(1995) 4153; Nucl. Phys. {\bf A596} (1996) 515.
\bibitem{KSW96} D. Kaplan, M. Savage and M. Wise, Nucl. Phys. {\bf
B478} (1996) 629. 
\bibitem{KSW98} D. Kaplan, M. Savage and M. Wise, Phys. Lett. {\bf
B424} (1998) 390; Nucl. Phys. {\bf B534} (1998) 329.
\bibitem{TH98} T. Cohen and J. Hansen, Phys. Lett. {\bf B440} (1998)
233; Phys. Rev. {\bf C59} (1999) 13; 
Phys. Rev. {\bf C59} (1999) 3047; nucl-th/9908049.
\bibitem{EGM98} E. Epelbaum, W. Gl\"ockle and U. Mei\ss ner,
Nucl. Phys. {\bf A637} (1998) 107; Nucl. Phys. {\bf A671}
(2000) 295.
\bibitem{FTT99} T. Frederico, V.S. Tim\'oteo and L. Tomio,
Nucl. Phys. {\bf A653} (1999) 209.
\bibitem{PRS99} D. R. Phillips, G. Rupak, M. Savage, Phys. Lett. {\bf
B473} (2000) 209.
\bibitem{FMS00} S. Fleming, T. Mehen and I. Stewart, Nucl. Phys. {\bf
A677} (2000) 313; Phys. Rev. {\bf C61} (2000) 044005.
\bibitem{Lu00} M. Lutz, Nucl. Phys. {\bf A677} (2000) 241.
\bibitem{GJ01} J. Gegelia, Phys. Lett. {\bf B463} (1999) 133; 
J. Gegelia and G. Japaridze, Phys. Lett. {\bf B517}
(2001) 476.
\bibitem{Be01} S. Beane, P. Bedaque, M. Savage , U. van
Kolck,  Nucl.Phys. {\bf A700} (2002) 377. 
\bibitem{ES01} D. Eiras and J. Soto, nucl-th/0107009.
\bibitem{BB02} T. Barford and M.C. Birse, hep-ph/0206146.
\bibitem{EM02} D.R. Entem and R. Machleidt, Phys. Lett. {\bf B524}
(2002) 93.
\bibitem{Ol02} J. Oller, nucl-th/0207086.
\bibitem{ej99} J. Nieves and E. Ruiz Arriola, Phys. Lett. {\bf B455}
(1999) 30; Nucl. Phys. {\bf A679} (2000) 57.
\bibitem{ej01} J. Nieves and E. Ruiz Arriola, Phys. Rev. {\bf D64}
(2001) 116008; C. Garc\'\i a-Recio, 
J. Nieves, E. Ruiz Arriola and M. J. Vicente-Vacas, Phys. Rev. {\bf D67}
(2003) 076009.
\bibitem{mb02} T. Inoue, E. Oset and M.J. Vicente-Vacas,
Phys. Rev. {\bf C65} (2002) 035204.

\bibitem{O97} J.A. Oller and E. Oset, Nucl. Phys. {\bf A620} (1997)
438; J.A. Oller, E. Oset and J.R. Pel\'aez,
Phys. Rev. Lett. {\bf 80} (1998) 3452; Phys. Rev. {\bf
D59} (1991) 074001.


\bibitem{OR98}  E. Oset and A. Ramos,
Nucl. Phys. {\bf A 635} (1998) 99.


\bibitem{GL84} J. Gasser and H. Leutwyler, Ann. of Phys., NY {\bf 158}
(1984) 142.
\bibitem{NJ93} V. Stocks, R.Klomp, M. Rentmeester and J. de Swart,
Phys. Rev. {\bf C48} (1993) 792.



\bibitem{jn03} J. Nieves in preparation.

\end{thebibliography}
\end{document}